\documentclass[conference]{IEEEtran}
\usepackage{geometry}
\usepackage[
style = ieee,
sorting=none,
backend=bibtex]{biblatex}
\usepackage{amsmath,amssymb,amsfonts}
\usepackage{algorithmic}
\usepackage{graphicx}
\usepackage{textcomp}
\usepackage{xcolor}
\usepackage{hyperref}
\hypersetup{breaklinks=true}
\def\BibTeX{{\rm B\kern-.05em{\sc i\kern-.025em b}\kern-.08em
    T\kern-.1667em\lower.7ex\hbox{E}\kern-.125emX}}
\begin{document}
\title{\vspace{0.25in}A Review of the Operational Use of UAS in Public Safety Emergency Incidents
}

\author{\IEEEauthorblockN{Hunter M. Ray}
\IEEEauthorblockA{\textit{Department of Aerospace Engineering} \\
\textit{University of Colorado Boulder}\\
\textit{Boulder Emergency Squad}\\
Boulder, Colorado, USA \\
hunter.ray-1@colorado.edu}
\and
\IEEEauthorblockN{Capt. Ryan Singer}
\IEEEauthorblockA{\textit{Boulder Emergency Squad} \\
Boulder, Colorado, USA \\
ryansinger@boulderrescue.org}
\and
\IEEEauthorblockN{Nisar Ahmed}
\IEEEauthorblockA{\textit{Department of Aerospace Engineering} \\
\textit{University of Colorado Boulder}\\
Boulder, Colorado, USA \\
nisar.ahmed@colorado.edu}
}

\maketitle

\begin{abstract}
The domain of public safety in the form of search \& rescue, wildland firefighting, structure firefighting, and law enforcement operations have drawn great interest in the field of aerospace engineering, human-robot teaming, autonomous systems, and robotics. However, a divergence exists in the assumptions made in research and how state-of-the-art technologies may realistically transition into an operational capacity. To aid in the alignment between researchers, technologists, and end users, we aim to provide perspective on how small Uncrewed Aerial Systems (sUAS) have been applied in 114 real world incidents as part of a technical rescue team from 2016 to 2021. We highlight the main applications, integration, tasks, and challenges of employing UAS within five primary use cases including searches, evidence collection, SWAT, wildland firefighting, and structure firefighting. Within these use cases, key incidents are featured that provide perspective on the evolving and dynamic nature of UAS tasking during an operation. Finally, we highlight key technical directions for improving the utilization and efficiency of employing aerial technology in all emergency types.
\end{abstract}

\begin{IEEEkeywords}
Human-Robot Interaction, sUAS, UAS, Public Safety, Search and Rescue, Autonomous Systems
\end{IEEEkeywords}

\section{Introduction}
The increasing capability, decreasing costs, and ease of use of Uncrewed Aerial Systems (UAS) have led to the growing adoption of these aircraft by emergency services engaged in public safety. Public safety incidents include operations conducted under the broad banner of firefighters, emergency medical services (EMS), law enforcement, and other technical rescue services. The dynamic, uncertain environments inherent in the response to an emergency challenges the capabilities of these highly flexible platforms, which have continued to prove their worth in assisting first responders. These incidents, especially search and rescue, are also the subject of significant research within the field of autonomous systems \cite{BurksThesis}, human-machine interaction \cite{murphy2004human}, manned-unmanned teaming \cite{wang2011scalable}, controls \cite{liu2013robotic}, and human factors \cite{chen2017advances}. This is due to the fact that analyzing and applying systems to the field of emergency response parallels and compliments interactions found in other industries including defense, planetary exploration, medicine, and personal robotics. However, many researchers and engineers do not have a strong grasp of how robotic platforms, such as UAS, are currently employed within these environments, leading to a divergence between the research \& development of advanced capabilities and their realistic potential for integration with first responders. The objective of this paper is to provide context and awareness for how UAS have been integrated within a technical rescue team in Boulder County, Colorado over the past five years and the methods by which they have been employed on a range of emergency incidents. We hope that this information can be used to guide further research and aid in the implementation of advanced capabilities with end users. 

\begin{figure}[ht]
    \centerline{
    \includegraphics[width = 0.95\linewidth]{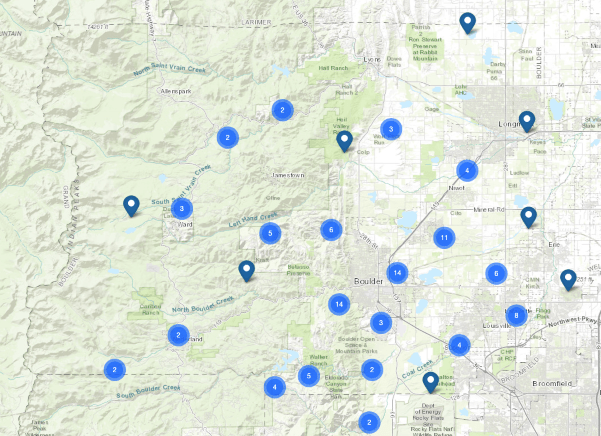}}
    \caption{Map of all analyzed calls performed in and near the diverse terrain found across Boulder County. Credit: D4H}
    \label{fig: Call_breakdown}
\end{figure}

This paper analyzes 114 calls completed by the Boulder Emergency Squad (BES) in which UAS were utilized from April 2016 to December 2021. Figure \ref{fig: Call_breakdown} displays the variety of locations and associated terrain where these calls took place. While other papers have covered specific implementations of UAS in one or two active disaster zones \cite{murphy_two_2016}\cite{surmann_deployment_2021}\cite{mehta_field_2020}\cite{murphy_use_2016}\cite{kruijff2012rescue} or field trials \cite{karma_use_2015} \cite{goodrich_towards_2009}, we present a unique overview of a wide range of deployments in real-world incidents. 

All of the incidents described here have been apart of a deployment with BES. BES is a 501.c.3 non-profit organization founded over 60 years ago for the preservation of life and property in Boulder County. BES provides a variety of technical rescue services to the County including dive rescue, swiftwater rescue, advanced extrication, rope rescue, wildland firefighting, urban search and rescue, UAS, and wide area search. This allows research conducted with BES to be evaluated over a wide variety of services and in both the rugged mountains and expansive plains found across the County, as shown in Figure \ref{fig: Call_breakdown}. Regarding UAS specific activities, BES provides UAS services for 21 different fire departments and 4 different law enforcement agencies in Boulder County. This includes deploying UAS in support of wildland fires, search and rescue operations, hazardous materials spills, SWAT and bomb related threats, and in the case of large-scale human or natural caused disasters. Currently, BES operates a fleet of 5+ aircraft with a variety of mission specific functions. Aircraft are operated by FAA Part 107 licensed remote pilots who receive continual training to improve their deployment of UAS and their effectiveness while on a mission.

We draw upon our own experiences as first responders, detailed logs, and interviews from fellow rescuers to synthesize the interactions found across diverse incidents. As researchers, we also provide a unique viewpoint on potential avenues for improving the integration, utilization, and effectiveness of UAS flight teams. From a regulatory and training standpoints, improved integration includes updated regulations and training to enable beyond visual line of sight operations and improved coordination with manned aircraft. From a research perspective, we note that current autonomous functions are insufficient to dynamic situations and future development necessitates a focus on making control more intuitive and flexible to reduce the number of operators required to field a single platform, as well as increase operator mobility and situational awareness.

The rest of the paper is organized as follows. Section II discusses how flight teams are integrated within the incident command structure, and the various roles that are currently required to field aerial platforms. Section III defines the specific hardware and vehicles used to support UAS activities on incidents. Section IV breaks down the surveyed incidents into five categories including search, evidence collection, special weapons and tactics (SWAT), wildland firefighting, and structure firefighting. Section V provides two detailed reports of notable incidents, including a wildland fire, and a missing person search. Section VI discusses opportunities for technological improvements of UAS flight teams. Finally, Section VII concludes the paper with primary takeaways.

\section{Team Structure}

Communication often presents the weakest link in any incident response. It is therefore vital that researchers understand systems of communication and coordination in public safety incidents to successfully develop technologies for responders. New technologies, such as autonomous capabilities, need to seamlessly integrate, as opposed to heavily modifying, the command structures to be successfully adopted. It’s important that proposed functions should address where the feature would be integrated, and if not, how the organizational structure should be modified to account for it. The following describes standards for the Incident Command System (ICS) defined by the National Emergency Management Systems \cite{ICS_manual}, as well as UAS specific leadership structures devised and implemented by BES.

\begin{figure}[htbp]
    \centerline{
    \includegraphics[width = 0.8\linewidth]{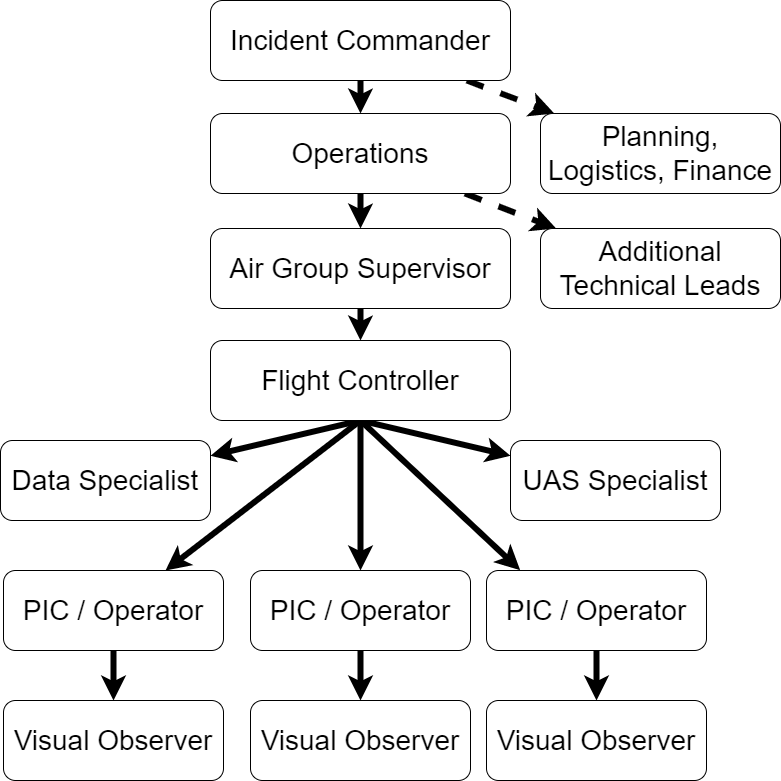}}
    \caption{Command hierarchy implemented during UAS incidents, with flight team components highlighted.}
    \label{fig: ICS_structure}
\end{figure}

\subsection{The Place of UAS within the Incident Command System}

Responders assisting in the direction and pilotage of UAS, and the dissemination of their data products, integrate within the command structure established during an incident. On emergency incidents, personnel establish themselves in accordance with the guidelines of the Incident Command System (ICS), which provides a modular, top-down framework detailing how incident response can be organized at all levels of government. An example ICS structure with a UAS focus is shown in Figure \ref{fig: ICS_structure}. Different functions can be stood-up by multiple people or single individuals as the incident evolves. At the top of the structure sits the Incident Commander (IC), who maintains authority of the organization, sets the overall objectives, and defines the priorities. Below the IC, there can be up to four specialized sections including: Operations, which manages the tactical response; Planning, which collects and disseminates intelligence, and plans operations; Logistics, which provides all necessary services to incident personnel; and Finance, which assists in financial aspects of the incident. The ICS structure allows for modular expansion or contraction of each of these sections' responsibilities as the IC deems fit for the particular response. On smaller incidents, such as a small house fire, the IC may fulfill all of the sections’ responsibilities themselves. Conversely, a widespread wildland fire may have multiple ICs and section Chiefs in all branches who alternate to maintain command throughout the incident.

UAS activities take place within the Operations section as part of the Air Group. Within Operations, each specialty will have an established functional supervisor. For example, a complex water rescue may have a Water Group Supervisor, Ropes Supervisor, Medical Supervisor, and Air Group Supervisor who all report to the Operations Chief. The Air Group Supervisor coordinates the aerial activities, such as coordinating helicopter evacuations or UAS operations, within the context of the incident by collaborating with the Operations Chief and other specialty supervisors. They define and prioritize the operational plan, coordinate with other agencies, obtain airspace authorizations, and ensure the safety of involved aircraft. 

The Flight Controller takes assignments from the Air Group Supervisor and disseminates plans to the respective flight team members. The Flight Controller orchestrates the details of the operation by coordinating available air assets within the defined airspace, including defining landings zones, and managing takeoffs and landings. For example, if the Supervisor assigns the task of maintaining visual coverage of a portion of a burning building, the Flight Controller will define the number of aircraft required and coordinate their hand-off on station as batteries are swapped and charged. 

Reporting to the Flight Controller include the individual flight teams, a UAS Specialist, and Data Specialist. A flight team is led by the Pilot-in-Command (PIC), who must maintain relevant flight qualifications, such as a FAA Part 107 certification. They may be assisted by a visual observer, who keeps the aircraft within line of sight at all times (this can also be done by the PIC themselves), and a scribe, who maintains a record of takeoff \& landing times and battery levels. Recording battery levels allows all resources to be accounted for, which is especially important on extended incidents, and for long-term maintenance purposes. A UAS Specialist assists in changing aircraft configurations, marking landing zones, and maintaining connectivity of data products as they’re disseminated across the command structure. Finally, a data specialist reviews real time data, processes imagery, and stores relevant flight data. While in theory every role could be filled by a specific individual, the nature of the responsibilities allow an individual to concurrently fill multiple roles. 

\begin{figure}[ht]
    \centerline{
    \includegraphics[width = 0.9\linewidth]{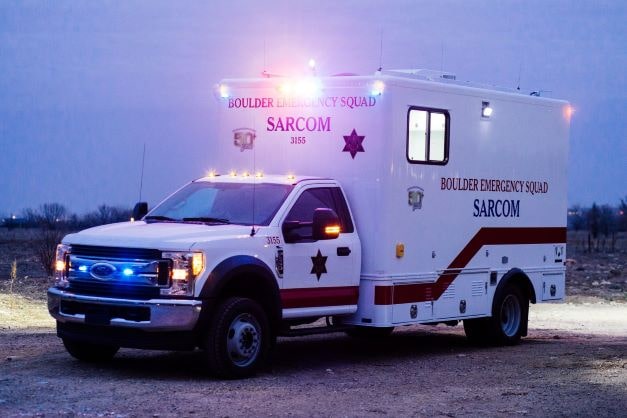}}
    \caption{The SARCOM mobile command vehicle is based on an all-terrain chassis that enables access to remote corners of Boulder County.}
    \label{fig: SARCOM}
\end{figure}
\begin{figure}[ht]
    \centerline{
    \includegraphics[width = 0.9\linewidth]{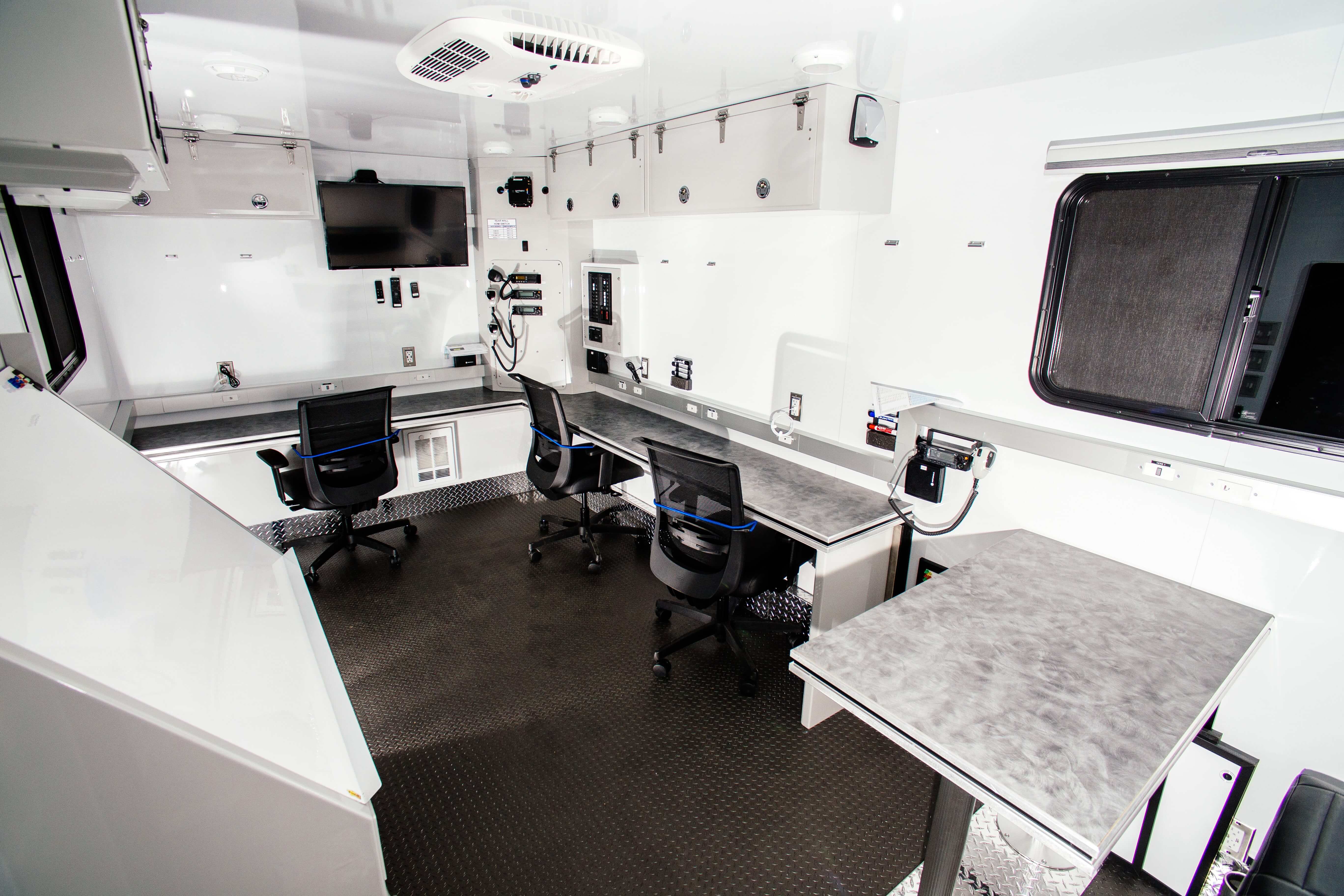}}
    \caption{The interior of the SARCOM vehicle provides an integrated center for directing operations, charging aircraft batteries, and disseminating data products across the command chain.}
    \label{fig: SARCOM_inside}
\end{figure}

\section{Hardware}
BES started providing UAS service in 2016 with a DJI Phantom 2 UAS. Since then, the fleet has grown to over five aircraft from the DJI family including Mavic, Mavic Advanced, and M210.  Purchased in 2017, a DJI M210 increased the squad’s utilization of UAS in adverse conditions, such as higher sustained winds of up to 32.8ft/s (10m/s), at night with a FLIR camera, and new situations with increased payload capacity. This aircraft was recently configured with a payload delivery system, which has yet to be utilized in an operational environment. The Mavic series of aircraft boast a top speed of 45mph (72kph), with a flight time of 25-30 minutes depending on its configuration and how the aircraft is flown. The Mavic Advanced have a listed range of ~6 miles (10km), but they are rarely flown at this distance in practice. These aircraft also include an IR camera, which can surpass the capabilities of the older M-210 hardware. Looking forward, the primary considerations in aircraft selection included the ability to meet mission requirements, ease of use, logistical support, and cost. A future enterprise platform will likely be a hybrid-quadcopter design, whose long endurance and familiar flight characteristics enable it to expand upon current operational needs.

Maintaining a fleet of aircraft allows platforms to be spread across more vehicles and ensures that UAS are available on all incidents. However, more aircraft require the maintenance and repair of sufficient spare batteries for long operations. The addition of rapid battery chargers that can simultaneously charge four batteries to 90\% in 30 minutes allows field teams to quickly swap battery packs prior to retasking.

Inaugurated in July 2017, the BES Search and Rescue Communications (SARCOM) mobile command vehicle enables integrated and enduring UAS services to a variety of incident types. Shown in Figure \ref{fig: SARCOM}, it is built off a Ford F-550 XLT pickup chassis, allowing the vehicle to be rugged enough to reach incident locations along the narrow, unpaved roads found across much of the County. The walk-in passenger box, shown in Figure \ref{fig: SARCOM_inside}, contains a variety of computers, monitors, and a printer to aid in the planning and command of incidents. It also carries a complement of communications equipment including VHF radio, 800 Mhz radio, WiFi antennas, VHF aircraft band radios, and LTE connectivity. Working alongside UAS, the SARCOM enables continuous A/V streaming and data dissemination across the command chain. Additionally, the 7KW onboard generator allows for concurrent charging of UAS batteries, video streaming \& display, and indoor climate control.

\section{Incident Types}
\begin{figure}[ht]
    \centerline{
    \includegraphics[width = 0.9\linewidth]{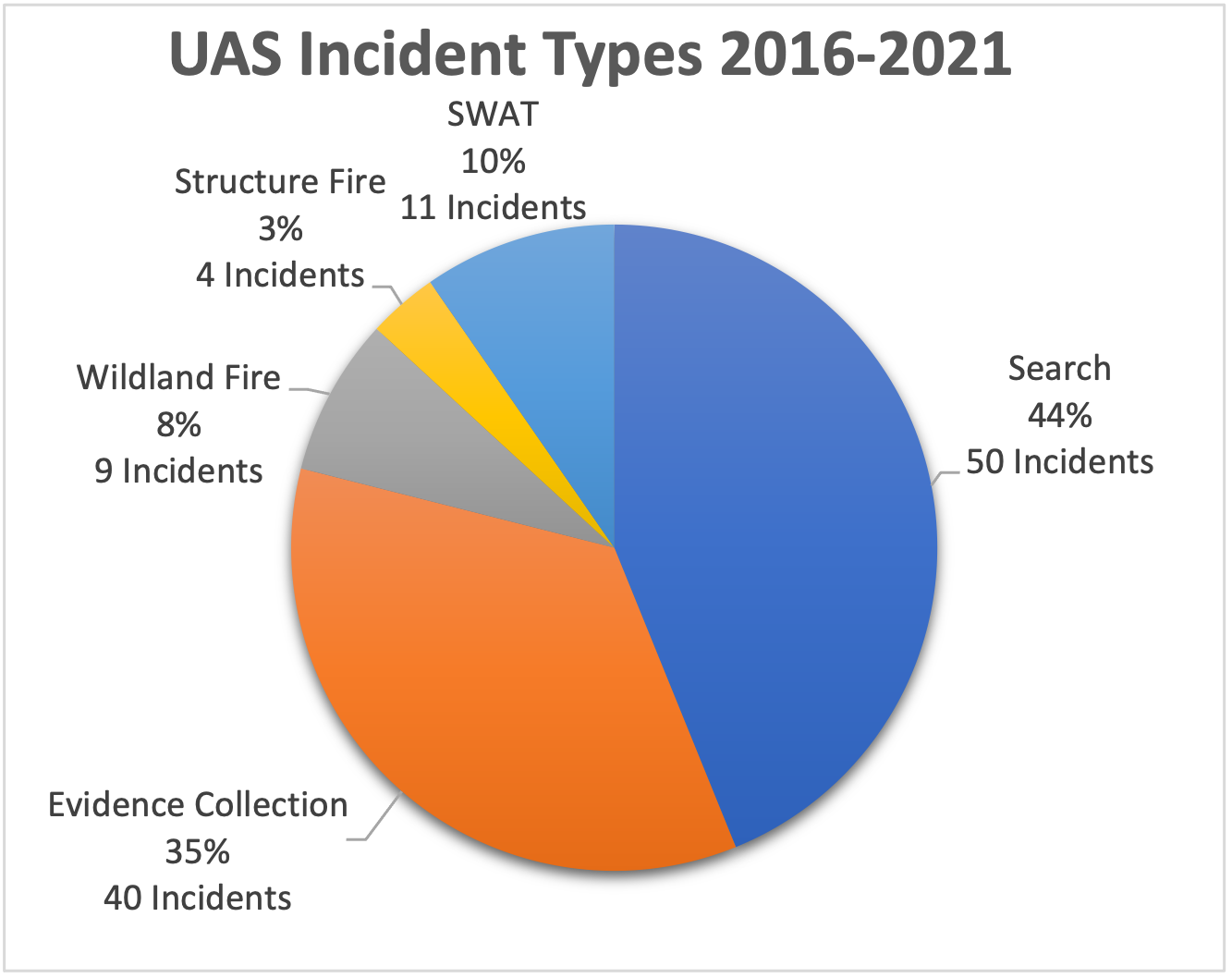}}
    \caption{Breakdown of reviewed UAS calls by incident type.}
    \label{fig: UAS_call_breakdown}
\end{figure}

UAS provides a flexible platform to assist in a wide variety of incident types. The following section provides a high-level overview of the five specific types of incidents where UAS have been utilized. Figure \ref{fig: UAS_call_breakdown} details the 114 UAS-involved incidents over the last five years from 2016-2021, which have been broken down into Search, Evidence Collection, Special Weapons and Tactics (SWAT), Wildland Fire, and Structure Fire. Each of these instances incorporates unique tactics and requirements for integration of the flight teams and dissemination of UAS data products across the command chain.  

\begin{figure}[ht]
    \centerline{
    \includegraphics[width = 0.9\linewidth]{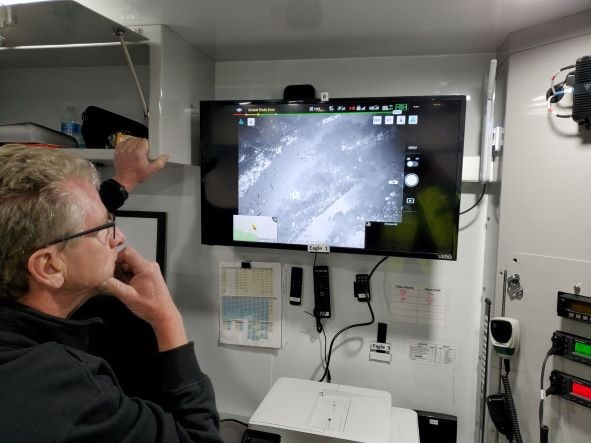}}
    \caption{During searches, one person is often tasked with supervising the video feed for possible evidence, sometimes inside the SARCOM vehicle.}
    \label{fig: watching_drones}
\end{figure}

Search incidents involve the localization of missing persons or evidence in urban, sub-urban, and natural environments. During a search, flight teams are assigned segments of the operational area to cover with their aircraft to look for the target of interest. These areas are usually divided along terrestrial boundaries such as fence lines, power lines, ridges, bodies of water, or forests. UAS operators and their flight team are tasked with monitoring video feeds, looking for the potential target of interest or possible clues leading to their discovery, such as clothing. The diverse types of possible evidence, significant obstructions, such as looking through trees, and limited detail available over remote video feeds complicate the automation of this task. As an area is searched, aircraft will be able to fly anywhere from a few hundred feet to several miles of the pilot, if line-of-sight and weather conditions permit. However, flying at a long distance and low altitude over rugged terrain presents a significant challenge, especially when a pilot must concurrently scan the video and pilot the aircraft. In these situations, the UAS data specialist assists the pilot in searching the video feeds, as shown in Figure \ref{fig: watching_drones}. Aircraft noise can also play a role in mission success as one missing person reported that the aircraft noise roused them sufficiently to call for help. Alternatively, while assisting law enforcement on a fugitive search at night, aircraft noise alerted the suspect to the ongoing search, potentially leading to their escape. In a separate incident, the noise was reported to have pinned down pursued subjects in a ditch as they believed they could easily be seen, which led to their arrest when located by a K9 team. Automated functions available on DJI hardware, including lawnmower search patterns and waypoint navigation, are rarely used during searches due to their limited applicability and long setup time. The strict geometrical boundaries required for lawnmower patterns doesn’t effectively meet the needs of the incident as areas assigned to flight teams often include diverse terrain types, some of which may require more direct investigation than others. For example, a person is more likely to be sheltering in a ditch than the nearby field, and these different environments require unique needs in terms of vehicle speed and perspectives to reliably clear them. Furthermore, changing terrain heights in the mountains leads to the need for a dynamic above ground level (AGL) flight altitude, which current automation cannot effectively support. 

\begin{figure}[ht]
    \centerline{
    \includegraphics[width = 0.9\linewidth]{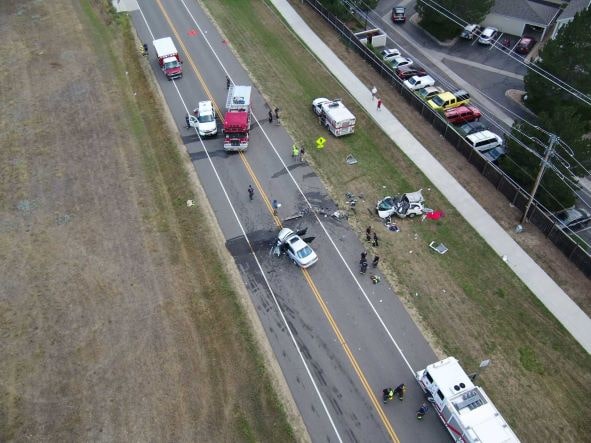}}
    \caption{UAS provide a unique perspective and critical evidence when reconstructing accident scenes.}
    \label{fig: Evidence_Collection}
\end{figure}

Evidence collection involves taking photographs or videos of a crime scene, an inactive fire location, or accident. Utilization of UAS provides multiple perspectives to improve context, which can be especially helpful during investigations. One or two flights by the aircraft are often sufficient to collect the required evidence, gathering an overhead view as shown in Figure \ref{fig: Evidence_Collection}. In these operations, autonomous modes are often utilized to ensure full area coverage and aid in photogrammetric reconstructions. Recognizing the utility of UAS in these events, law enforcement have started to purchase and utilize their own aircraft to assist in a larger number of their investigations. Aerial imagery of fire locations can also be helpful in understanding fire progression, especially in the initial investigation of wildland fire origins. If the incident involves a crime, care must be taken to ensure chain of custody of the SD card from pilot-in-command to law enforcement. 

BES maintains a long history of working with law enforcement and SWAT teams on investigations, raids, and other types of security operations. In particular, UAS visual and IR cameras have become increasingly valuable in assisting officers in dangerous incidents including active shooters, barricaded subjects, warrant service, and VIP security. The primary mission of UAS during these encounters involves the improvement of situational awareness, which can manifest in a number of ways. If the hot zone is a known location, multiple aircraft can be used to provide continuous 360° coverage of the area. During a tragic active shooter incident at a local supermarket in Boulder, at least four aircraft from multiple agencies were used to maintain video on all building sides, during which temporary requests for detailed investigations were also made via the command structure. UAS have also been used to follow advancing officers and armored vehicles, providing extra visibility for the surrounding areas and over obstacles. In most instances, video products are relayed to the SARCOM vehicle for local analysis by commanding officers and dissemination, such as to SWAT officers in the hot zone. In one case, a BES flight team rode inside an armored vehicle to safely approach the target location and perform reconnaissance, however, the team was later required to relocate to properly share video with command. Key challenges for flying UAS within these incidents include maintaining LOS and aircraft control while the aircraft is in the hot zone, prompt relay of video products to officers on the ground, and accurate direction of aircraft by command or ground teams. 
 
UAS employed in wildland firefighting provide value in detecting spot fires, and aiding firefighters in situational awareness in these highly dynamic events. Boulder County is particularly vulnerable to destructive wildfires, notably the Calwood Fire of October 2020, and the Marshall Fire of December 2021. However, UAS are more often utilized on smaller, developing fires prior to the engagement with larger manned aircraft, or after the climax of the fire's progression for targeted extinguishing. This is particularly the case with the investigation, detection, and assistance in mitigation of spot fires caused by lightning strikes or traveling embers from an ongoing fire. Due to the challenging terrain, nature of the task, and weather conditions, autonomous functions have never been used in these incidents. The primary challenges associated with utilizing UAS to assist in wildfires include coordination with manned aircraft, and wind. Manned aircraft are often employed to drop water and fire retardant, which requires them to fly at low altitudes. Previous encounters with unauthorized UAS by manned aircraft pilots across the country have caused significant hesitancy towards the collaborative utilization of UAS and manned aircraft. Additionally, large fires are often fueled by high winds, which stymie the deployment of UAS, such as during the Marshall Fire where 100mph gusts grounded all aircraft \cite{MarshallFire}. Key components to enabling a safe response to a wildfire involve the emplacement of lookouts, communications between personnel about developing conditions, awareness of multiple escape routes, and knowledge of safety zones, summed up in the often repeated acronym “LCES” \cite{LCES_basics}. UAS are poised to potentially play a role in all of these components by embedding flight teams within ground crews, and employing aircraft as repeaters for improved communications in mountainous terrain. 

The final type of incidents in which BES deployed UAS involved structure fires, such as the burning of homes, businesses, and barns. Firefighters use UAS equipped with IR cameras to detect hot zones within the remains of the structure, and live power lines, whose detection helps firefighters direct their suppression efforts. IR capability could be used to detect persons still trapped within the structure, however, this has yet to be performed BES in an active incident in Boulder County. 

\section{Detailed Incident Reports}
This section presents detailed after-action reports on certain notable incidents. These narratives provide more specific context about the interactions between flight teams, aircraft, supervisors, and ground personnel. The objective in presenting these summaries is to demonstrate the dynamic and evolving nature of incident response, as well as the specific value and challenges with employing UAS.

\begin{figure}[ht]
    \centerline{
    \includegraphics[width = 0.9\linewidth]{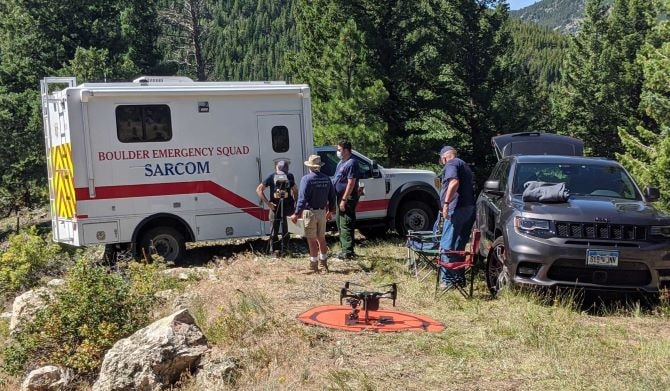}}
    \caption{Staging area and landing zone during the Coal Creek Wildfire Incident.}
    \label{fig: GranitePeak Images 1}
\end{figure}

\begin{figure}[ht]
    \centerline{
    \includegraphics[width = 0.9\linewidth]{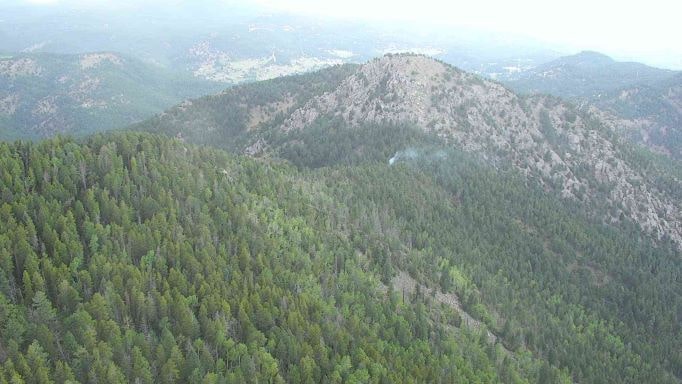}}
    \caption{Narrow plume of smoke captured by UAS during the Coal Creek Wildfire Incident}
    \label{fig: GranitePeak Images 2}
\end{figure}


\subsection{Wildland Fire, August 6, 2020}
On August 6, 2020, BES was dispatched to the Coal Creek area of Boulder County to assist a crew of wildland firefighters with an aerial survey of two spot fires caused by lightning strikes over a remote, mountainous area. Upon arrival to the staging area shown in Figure \ref{fig: GranitePeak Images 1}, a pilot was tasked to use their DJI Mavic UAS as a lookout for the hand crews digging fire lines around the first fire. Concurrently, the M210 was sent to a second spot fire over 1 mile away from the incident command post. This aircraft also acted as a lookout for evolving fire situations to the single firefighter who was on the ground at the time while additional personnel accessed the remote location. During the course of these events, a third lightning strike was reported by crews on the fire line and the M210 was re-tasked from its initial assignment to investigate the report. This aircraft located the smoke and growing 10ft by 10ft fire area, shown in Figure \ref{fig: GranitePeak Images 2}. Coordinates and imagery were provided to the IC, who was able to evaluate the severity of the fire and effectively task additional resources. As these crews mobilized, a nearby house and best access route for firefighters were spotted with the help of the aircraft. After being briefly grounded by a passing rain storm, the aircraft re-evaluated the fire, which had grown to a size of 50ft by 20ft. Flight teams coordinated with arriving firefighters by using the aircraft as a beacon to direct them to the best access route, where they were able to quickly mitigate the fire’s spread. As night fell, UAS were also used to direct firefighters to egress routes. The next day, BES was dispatched to the same area to confirm the extinguishment of the previous day’s fires. Flight teams were tasked to utilize IR cameras to review the heat signatures of the various fires, during which crews had to be dispatched for further mitigation with water packs. UAS also scouted the local area to ensure no additional fires had been started. 

This incident provides a case study in the substantial value that UAS add to a remote wildfire incident, and a few challenges that were faced merit further discussion. First, the ground crews and aircraft had difficulty locating each other. Crews eventually used reflective surfaces to capture the attention of the aircraft, however this points to the need for more effective methods for collaboration between aerial and mobile ground teams. Second, the location of the incident command post provided an opportunistic viewpoint from which to fly aircraft at long distances and remain within pilots’ line of sight. However, these expansive vantage points cannot be counted on and methods for flying beyond visual line of sight or integrating pilots within firefighter crews should be further explored. 

\subsection{Search for Missing Person, November 6, 2020}
\begin{figure*}[ht]
    \centerline{
    \includegraphics[width =0.95\linewidth]{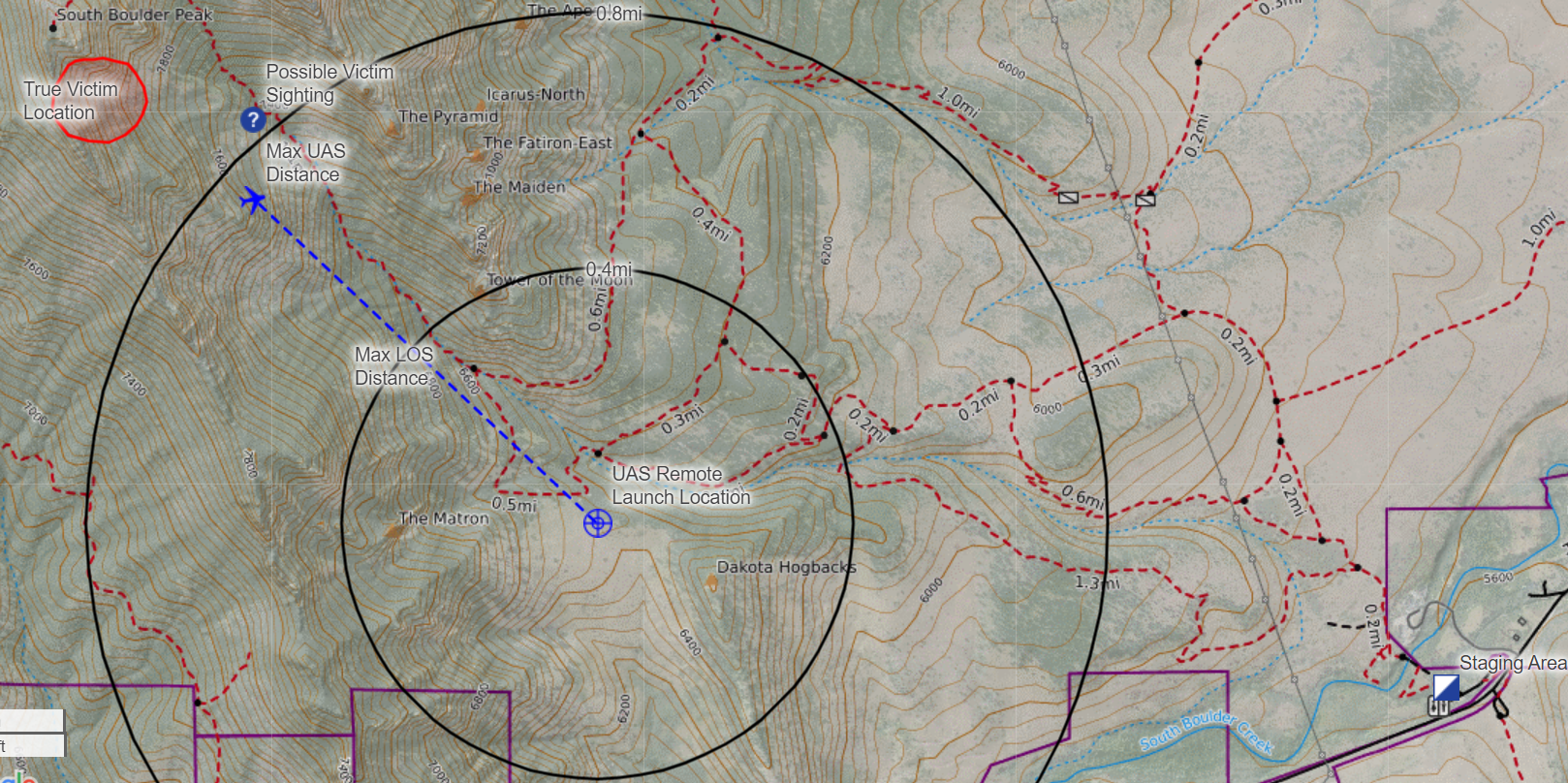}}
    \caption{A map of the South Boulder Mesa, which was searched during the November 6 incident. The incident command post and staging area is at the lower right of the map. The victim's shoe was found by a park ranger near the top of South Boulder Peak at the upper left. Credit: SARTopo}
    \label{fig: south Boulder map}
    \vspace{-8pt}
\end{figure*}
In the evening of November 6, 2020 at 19:11, BES was dispatched to the South Boulder Mesa trailhead to search for a missing, and possibly suicidal, hiker. The challenging location, perceived victim mental state, and lack of sunlight created additional urgency within this incident. This area of the county sits at the intersection of the high plains and the foothills of the Rocky Mountains, and is characterized by steep and sudden changes in elevation. Within a horizontal distance of 0.7 miles (1.1km) the mountains in this area rise by over 2000ft (600m). UAS were requested to search inaccessible locations within and around the rocky crags and gullies found across the mountain face. 

Within a few hours of the incident, whose layout is described in Figure \ref{fig: south Boulder map}, a Park Ranger found the victim's shoe on the top of South Boulder Peak. However, the 2.5 mile (3.9km) distance from the trailhead to mountain peak created significant challenges towards maintaining communications and LOS with the aircraft, which required the deployment of mobile flight teams onto mountain trails. The only platforms currently in service by the rescue team at this time was the DJI Mavic 210, which, due to its bulk and size, was not easily portable up narrow trails. A utility task vehicle (UTV) was used to shuttle a flight team up a wide path to a remote launch location that was considered to have clear LOS towards the peak. To direct search efforts, contact with the ranger was attempted and finally made after she shined her flashlight in its general direction. However, the aircraft was still approximately 2000ft (600m) away. While this distance could have been effective for searching in the daylight with the aircraft's 30x optical zoom camera, the FLIR camera available aboard the aircraft couldn't accurately resolve detail in the cliffs at such a distance. During the search, a possible victim was sighted near the aircraft, however could not be confirmed from the air. Being at an altitude of approximately 1600ft (500m) above the launch point, software limitations prevented any further altitude increases, despite the peak being 430ft (130m) higher. The lack of effective launch positions, potential victim sighting by the ranger, and incoming inclement weather led to the conclusion of aerial operations. Unfortunately, the victim succumbed to their injuries.  

While UAS should have added value in this type of incident, a number of challenges hindered more effective application. Particularly notable are the difficulties faced with employing UAS in remote, mountainous environments without clear access or lines of sight to the area of operations. The large horizontal and vertical distances from landing zone to the search area also burdened the battery capacity, leaving the aircraft little time on station once it reached the search area. Improving flight teams' mobility, with respect to equipment portability and control of the aircraft, could have enabled a landing zone closer to the search area, or allowed the aircraft to be controlled from more rugged areas with better lines of sight throughout the search.

\section{Discussion}

As we’ve shown, UAS provide substantial value in assisting first responders in a variety of emergency situations. We conclude with a discussion of the perceived limitations in current technology, potential avenues for further exploration, as well as the integration of UAS in novel emergency types. UAS technology and their integration with incident responses are currently limited by the requirement for line-of-sight flying, the lack of integration with manned aircraft, the static and focused nature of current piloting modes, and the lack of reliable computer vision. 

Pilots operating under FAA Part 107 certification require their aircraft to be in their line of sight at all times. While the technology can usually be operated safely beyond visual line-of-sight (BVLOS), the lack of effective vantage points, such as that found during the Coal Creek Fires, can substantially mitigate the effective range of UAS in most environments. However, BVLOS operations can also challenge the connectivity of operators to the UAS, necessitating resilient and autonomous contingency functions. Regulators working with UAS operators should develop training and update regulations to enable safe BVLOS flying, including drawing from the experience of military operators. Further work should also consider how UAS should be integrated with manned aircraft during incidents, especially with respect to wildland firefighting. Utilizing GPS to localize UAS positions on incidents enables supervisors to coordinate and delineate airspace between platforms with sufficient horizontal and vertical safety margins. Ongoing work in automated collision avoidance systems for UAS, such as ACAS-Xu \cite{owen2019acas}, augment GPS localization with other collision avoidance systems, such as ADS-B, can also assist pilots with ensuring the safety of the involved aircraft.  

A major hindrance to more widespread utilization and effectiveness of UAS in emergency responses is the burden placed upon the pilot for maintaining control and directing the attention of the aircraft’s sensors. Pilots currently operate fielded UAS using sticks for flight control and a series of wheels, or a touchscreen to direct the camera’s attention. This type of control requires the operator to be completely stationary to focus on the task at hand, limiting their situational awareness and mobility. To support the pilot in their task, new roles, including the flight controller and data specialist, need to be created to ensure that their activities are well coordinated and their information is properly disseminated and analyzed. The multiple personnel required to support even a single aircraft highlights the potential challenges when considering the integration of multiple autonomous aircraft (i.e. swarms) for these types of incidents. As we’ve shown, UAS are utilized in a wide range of dynamic situations where aircraft can be given diverse and changing task assignments. Development of autonomous functions should focus on enabling pilots to communicate intentions in an intuitive manner irrespective of the particular nature of the incident. Integration of new technologies within these environments can apply co-active design principles that consider the interdependent nature of specific tasks and respectively leverage and compliment the rescuer and platform capabilities \cite{johnson2014coactive}. Autonomous functions such as optimal path planning for search \& rescue can only be achieved if all information can be effectively quantified for the algorithm, which necessitates the burdensome incorporation of contextual information from a particular incident. The objective of autonomous modes should not be to obviate the need for a rescuer, but instead to take over basic piloting tasks to improve an operator’s mobility and situational awareness, and reduce their workload. Enabling the intuitive direction of UAS in a manner that integrates the aircraft’s capabilities within the team presents the greatest opportunity for enabling flight teams to respond to and adapt to dynamic situations. Methods such as \cite{julian2019distributed}, which employ a POMDP algorithm aboard a UAS to track wildfire growth, are promising, but need methods to incorporate larger contextual information and constraints provided by operators to have a chance of being employed in active incidents. 

To a limited capacity, some autonomous functions are employed by operators on all incidents, and are noted for their integration within the primary ``manual" piloting mode. These can be broken up into functions for information synthesis and contingency flight modes. With respect to information synthesis, particularly notable is a battery life estimate, which take into account the current distance from the landing zone to the aircraft's current position to give an estimated remaining flight time. The associated metrics and audio warning features help the operator stay aware of their aircraft's status and the aircraft can even take control when the battery is at critical levels. Additional autonomous functions that synthesize flight information and detect upcoming failures, such as component fault detection \cite{Baskaya2017FaultDetection}, can be readily integrated into operational systems as they can be easily ignored or utilized by pilots without issue. The sole contingency flight mode available on DJI hardware is a return-to-home function, which is triggered when the battery is critically low, the aircraft suffers from a connection failure, or at the operator's request. Therefore, setting the return-to-home altitude is a critical component within the pre-flight checklist. It was observed that this mode was activated by pilots on a number of incidents even when the aircraft was operating effectively to aid in returning to the landing zone. Incorporating more sophisticated contingency flight modes that incorporate active obstacle avoidance and dynamic terrain will be critical towards enabling reliable BVLOS operations. However, ensuring that operators reliably trust the performance of these flight modes is another important aspect of their integration.

\begin{figure}[ht]
    \centerline{
    \includegraphics[width = 0.9\linewidth]{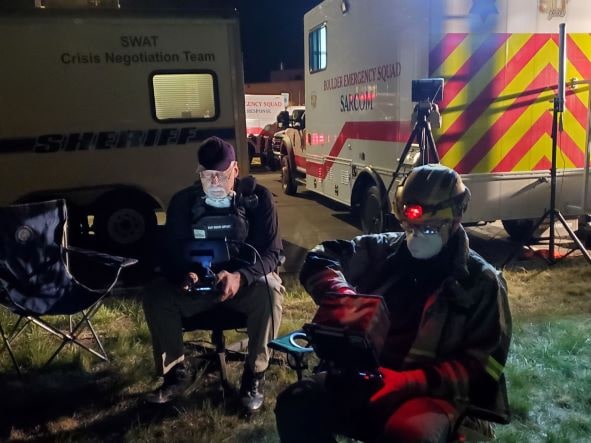}}
    \caption{Current methods of UAS pilotage require static pilots fully engaged in their tasks.}
    \label{fig: Pilots}
\end{figure}

Incorporating computer vision into video stream analysis of UAS derived video products presents another opportunity for further research. During a search with UAS, a single person may be required to supervise a video feed to check for evidence of the missing individual. Computer vision cannot be entirely relied upon due to the challenging perspectives required in most operations, and myriad of objects that may be desired. However, a limited application of computer vision could query an operator for investigation of perceived anomalies in IR signatures, colors, patterns, or shapes.

The flexible nature of UAS platforms allows potential for application in novel regimes and implementations. For example, a swiftwater search and rescue operation may entail a challenging exploration of a creek or riverway for possible evidence of the missing persons. While heavy undergrowth on the banks and dangerous rapids can present significant obstacles for ground and water teams, these incidents present a ripe opportunity for UAS to be leveraged for the search. UAS could also be used to deliver supplies and aid to stranded people, assist in avalanche rescue, and act as communication relays in remote areas. Additionally, novel modalities of integration could include deployable fixed-wing aircraft that can launch from a centralized location to provide earlier incident evaluations, similar to that being done by the Chula Vista Police department with quadcopter UAS\cite{ChulaVistaPolice}. 

\section{Conclusion}

It’s clear that UAS are increasingly providing value to first responders on diverse incident types. This paper demonstrates the various methods and structures by which UAS assist first responders in diverse public safety emergency incidents. An overview of the command structure by which flights are coordinated was presented and call attention to the multiple personnel often required to manage a single airborne platform. We reviewed 114 incidents where UAS were utilized to assist with searches, evidence collection, wildland firefighting, structure firefighting, and SWAT operations. Notable incidents that merit detailed recounting were highlighted to provide perspective on the dynamic nature of a particular response. Finally, current technological and regulatory limitations were discussed, and recommendations for areas of additional study were provided. 

 The integration of UAS with first responders allows a unique perspective in the study of human-machine interaction and manned-unmanned teaming. In the near term, researchers should focus less on making highly sophisticated ``out-of-the-box" autonomy that immediately performs with minimal supervision, which leads to brittle and unreliable performance. Instead, UAS technology development should aim towards enabling lower-level decision-making and perception features that allow for varying levels of integrated autonomy, operator trust, and aircraft control with minimal training. Within incidents, human-centered coordination of operations remains key, since the context and dynamics of an incident can vary drastically within and between missions for the same set of vehicles and operators. Aligning the needs of these end users with ongoing research thrusts will enable greater utilization of developed capabilities in public safety emergencies and other applications of human-robot teaming in complex, dynamic, and uncertain environments.

\section*{Acknowledgment}
The authors would like to thank the members of BES for their helpful and insightful contributions.

\printbibliography

@manual{ICS_manual,
	title		= {National Incident Management System},
	organization	= {Federal Emergency Management Agency},
	edition	= {3rd},
	year		= {2017},
	month		= {10},
	url = {https://www.fema.gov/sites/default/files/2020-07/fema_nims_doctrine-2017.pdf}
}

@article{MarshallFire,
    author    = {Blair Miller},
    title     = {National Weather Service analysis identifies factors that caused Marshall Fire to spread so quickly},
    journal   = {The Denver Channel},
    year      = {2022},
    month    = {1},
    url = {https://www.thedenverchannel.com/news/marshall-fire/national-weather-service-analysis-identifies-factors-that-caused-marshall-fire-to-spread-so-quickly}
}

@misc{LCES_basics,
    author    = {Paul Gleason},
    title     = {Lookouts, Communication, Escape Routes and Safety Zones, ``{LCES}''},
    howpublished = {Online},
    year     = {1991},
    month    = {6},
    url = {https://www.nwcg.gov/sites/default/files/wfldp/docs/lces-gleason.pdf}
}

@inproceedings{owen2019acas,
  title={{ACAS} {X}u: Integrated collision avoidance and detect and avoid capability for {UAS}},
  author={Owen, Michael P and Panken, Adam and Moss, Robert and Alvarez, Luis and Leeper, Charles},
  booktitle={2019 IEEE/AIAA 38th Digital Avionics Systems Conference (DASC)},
  pages={1--10},
  year={2019},
  organization={IEEE}
}

@misc{ChulaVistaPolice,
    author ={Chula Vista Police Department},
    title     = {Drones as First Responder: The Future of Public Safety},
    howpublished = {Online},
    year     = {2022},
    url = {https://www.chulavistaca.gov/home/showpublisheddocument/19749/637084636048200000}
}

@article{julian2019distributed,
  title={Distributed wildfire surveillance with autonomous aircraft using deep reinforcement learning},
  author={Julian, Kyle D and Kochenderfer, Mykel J},
  journal={Journal of Guidance, Control, and Dynamics},
  volume={42},
  number={8},
  pages={1768--1778},
  year={2019},
  publisher={American Institute of Aeronautics and Astronautics}
}

@book{BurksThesis,
    title = {Active Collaborative Planning and Sensing in Human-Robot Teams},
    author = {Luke Burks},
    publisher = {University of Colorado Boulder},
    year = {2020},
    location = {Boulder, Colorado}
}

@article{murphy2004human,
  title={Human-robot interaction in rescue robotics},
  author={Murphy, Robin R},
  journal={IEEE Transactions on Systems, Man, and Cybernetics, Part C (Applications and Reviews)},
  volume={34},
  number={2},
  pages={138--153},
  year={2004},
  publisher={IEEE}
}

@article{liu2013robotic,
  title={Robotic urban search and rescue: A survey from the control perspective},
  author={Liu, Yugang and Nejat, Goldie},
  journal={Journal of Intelligent \& Robotic Systems},
  volume={72},
  number={2},
  pages={147--165},
  year={2013},
  publisher={Springer}
}

@inproceedings{chen2017advances,
  title={Advances in Human Factors in Robots and Unmanned Systems},
  author={Chen, Jessie},
  booktitle={AHFE International Conference on Human Factors in Robots and Unmanned Systems. Florida: AHFE},
  year={2017},
  organization={Springer}
}

@inproceedings{wang2011scalable,
  title={Scalable target detection for large robot teams},
  author={Wang, Huadong and Kolling, Andreas and Brooks, Nathan and Owens, Sean and Abedin, Shafiq and Scerri, Paul and Lee, Pei-ju and Chien, Shih-Yi and Lewis, Michael and Sycara, Katia},
  booktitle={Proceedings of the 6th international conference on Human-robot interaction},
  pages={363--370},
  year={2011}
}

@article{karma_use_2015,
	title = {Use of unmanned vehicles in search and rescue operations in forest fires: {Advantages} and limitations observed in a field trial},
	volume = {13},
	issn = {22124209},
	url = {https://linkinghub.elsevier.com/retrieve/pii/S2212420915300364},
	doi = {10.1016/j.ijdrr.2015.07.009},
	journal = {International Journal of Disaster Risk Reduction},
	author = {Karma, S. and Zorba, E. and Pallis, G.C. and Statheropoulos, G. and Balta, I. and Mikedi, K. and Vamvakari, J. and Pappa, A. and Chalaris, M. and Xanthopoulos, G. and Statheropoulos, M.},
	month = sep,
	year = {2015},
	pages = {307--312},
}

@article{goodrich_towards_2009,
	title = {Towards using {Unmanned} {Aerial} {Vehicles} ({UAVs}) in {Wilderness} {Search} and {Rescue}: {Lessons} from field trials},
	volume = {10},
	shorttitle = {Towards using {Unmanned} {Aerial} {Vehicles} ({UAVs}) in {Wilderness} {Search} and {Rescue}},
	url = {http://www.jbe-platform.com/content/journals/10.1075/is.10.3.08goo},
	doi = {10.1075/is.10.3.08goo},
	language = {en},
	number = {3},
	journal = {Interaction Studies. Social Behaviour and Communication in Biological and Artificial Systems},
	author = {Goodrich, Michael A. and Morse, Bryan S. and Engh, Cameron and Cooper, Joseph L. and Adams, Julie A.},
	month = dec,
	year = {2009},
	pages = {453--478},
}

@inproceedings{murphy_two_2016,
	title = {Two case studies and gaps analysis of flood assessment for emergency management with small unmanned aerial systems},
	doi = {10.1109/SSRR.2016.7784277},
	booktitle = {2016 {IEEE} {International} {Symposium} on {Safety}, {Security}, and {Rescue} {Robotics} ({SSRR})},
	author = {Murphy, Robin and Dufek, Jan and Sarmiento, Traci and Wilde, Grant and Xiao, Xuesu and Braun, Jeff and Mullen, Lachlan and Smith, Richard and Allred, Sam and Adams, Justin and Wright, Adam and Gingrich, Jess},
	month = {10},
	year = {2016},
	pages = {54--61},
}

@inproceedings{surmann_deployment_2021,
	title = {Deployment of {Aerial} {Robots} after a major fire of an industrial hall with hazardous substances, a report},
	doi = {10.1109/SSRR53300.2021.9597677},
	booktitle = {2021 {IEEE} {International} {Symposium} on {Safety}, {Security}, and {Rescue} {Robotics} ({SSRR})},
	author = {Surmann, Hartmut and Slomma, Dominik and Grobelny, Stefan and Grafe, Robert},
	month = {10},
	year = {2021},
	pages = {40--47},
}

@article{mehta_field_2020,
	title = {Field {Methods} to {Quantify} {Emergency} {Responder} {Fatigue}: {Lessons} {Learned} from {sUAS} {Deployment} at the 2018 {Kilauea} {Volcano} {Eruption}},
	volume = {8},
	url = {https://doi.org/10.1080/24725838.2020.1855272},
	number = {3},
	journal = {IISE Transactions on Occupational Ergonomics and Human Factors},
	author = {Mehta, Ranjana K and Nuamah, Joseph and Peres, S. Camille and Murphy, Robin R.},
	month = jul,
	year = {2020},
	pmid = {33241982},
	pages = {166--174},
}

@article{murphy_use_2016,
	title = {Use of a {Small} {Unmanned} {Aerial} {System} for the {SR}-530 {Mudslide} {Incident} near {Oso}, {Washington}},
	volume = {33},
	issn = {1556-4967},
	doi = {10.1002/rob.21586},
	language = {en},
	number = {4},
	journal = {Journal of Field Robotics},
	author = {Murphy, Robin R. and Duncan, Brittany A. and Collins, Tyler and Kendrick, Justin and Lohman, Patrick and Palmer, Tamara and Sanborn, Frank},
	year = {2016},
	pages = {476--488},
}

@inproceedings{kruijff2012rescue,
    title={Rescue robots at earthquake-hit {Mirandola}, {Italy}: A field report},
    author={Kruijff, Geert-Jan M and Pirri, Fiora and Gianni, Mario and Papadakis, Panagiotis and Pizzoli, Matia and Sinha, Arnab and Tretyakov, Viatcheslav and Linder, Thorsten and Pianese, Emanuele and Corrao, Salvatore and others},
    booktitle={2012 IEEE international symposium on safety, security, and rescue robotics (SSRR)},
    pages={1--8},
    year={2012},
    organization={IEEE}
}

@inproceedings{Baskaya2017FaultDetection,
    author={Baskaya, Elgiz and Bronz, Murat and Delahaye, Daniel},
    booktitle={2017 IEEE/AIAA 36th Digital Avionics Systems Conference (DASC)}, 
    title={Fault detection  \& diagnosis for small {UAV}s via machine learning}, 
    year={2017},
    volume={},
    number={},
    pages={1-6},
    doi={10.1109/DASC.2017.8102037}
}

@article{johnson2014coactive,
    title={Coactive design: Designing support for interdependence in joint activity},
    author={Johnson, Matthew and Bradshaw, Jeffrey M and Feltovich, Paul J and Jonker, Catholijn M and Van Riemsdijk, M Birna and Sierhuis, Maarten},
    journal={Journal of Human-Robot Interaction},
    volume={3},
    number={1},
    pages={43--69},
    year={2014},
    publisher={Journal of Human-Robot Interaction Steering Committee}
}

\end{document}